\documentclass[aps,prl,twocolumn,groupedaddress]{revtex4}

\usepackage{graphicx}

\begin{document}

\title{Transport criticality of the first-order Mott transition 
in a quasi-two-dimensional organic conductor, 
$\kappa$-(BEDT-TTF)$_{2}$Cu[N(CN)$_{2}$]Cl}

\author{F. Kagawa,$^{1}$ T. Itou,$^{2}$ K. Miyagawa,$^{1,2}$ and 
K. Kanoda$^{1,2}$}

\affiliation{
$^{1}$Department of Applied Physics, University of Tokyo, Bunkyo-ku, 
Tokyo 113-8656, Japan 
\\
$^{2}$CREST, Japan Science and Technology Corporation, Kawaguchi 
332-0012, Japan
\\}

\date{\today}

\begin{abstract}
	An organic Mott insulator, $\kappa$-(BEDT-TTF)$_{2}$Cu[N(CN)$_{2}$]Cl, 
was investigated by resistance measurements under continuously controllable 
He gas pressure. 
	The first-order Mott transition was demonstrated by observation of 
clear jump in the resistance variation against pressure. 
	Its critical endpoint at 38 K is featured by vanishing of the 
resistive jump and critical divergence in pressure derivative of 
resistance, $|\frac{1}{R}\frac{\partial R}{\partial P}|$ , which are 
consistent with the prediction of the dynamical mean field theory and have 
phenomenological correspondence with the liquid-gas transition. 
	The present results provide the experimental basis for physics of 
the Mott transition criticality. 

\end{abstract}



\maketitle

	The Mott transition is one of the metal-insulator transitions (MIT) 
which are representative phenomena in highly correlated electrons. 
	The family of quasi-two-dimensional layered organic conductors, 
$\kappa$-(BEDT-TTF)$_{2}$X, are model systems for the study of the Mott 
transition in two dimensions, where BEDT-TTF is 
bis(ethylenedithio)tetrathiafulvalene and X stands for various kinds of 
anions. 
	In the conducting layer, the BEDT-TTF dimers form anisotropic 
triangular lattice, and the dimer band is half-filled \cite{Ref1:Kanoda}. 
	The salt of X = Cu[N(CN)$_{2}$]Cl (denoted as $\kappa$-Cl hereafter) 
is an antiferromagnetic insulator (AFI) with a commensurate order at ambient 
pressure and thus is understood as a Mott insulator driven by the strong 
electron correlation \cite{Ref2:Miyagawa}. 
	When $\kappa$-Cl is pressurized, it becomes metallic and undergoes a 
superconducting transition at $T_{\rm SC}\sim$ 13 K under a pressure of 
30 MPa \cite{Ref3:Williams}. 
	It is considered that the pressure induces the Mott transition. 
	The pressure is quite effective to drive the Mott transition in 
organics, which are highly compressible. 
	Moreover, since they have no orbital degree of freedom, they give a 
prototype of the Mott transition. 
	These aspects make organics suitable for pursuing the fundamentals 
of the Mott transition both experimentally \cite{Ref1:Kanoda} and 
theoretically \cite{Ref4:McKenzie,Ref5:Kino,Ref6a:Merino}.

	The behavior near the MIT is a key to comprehension of the Mott 
transition. 
	However, since most of the previous experiments have been performed 
under the chemical or discrete pressure control 
\cite{Ref3:Williams,Ref6:Kawamoto,Ref7:Taniguchi}, the critical nature of 
Mott transition has remained to be seen. 
	Recently, ac susceptibility and NMR studies for $\kappa$-Cl under 
continuously controllable He gas pressure by Lefebvre \textit{et al.} have 
revealed 
(i) the first-order nature of the Mott transition, suggested 
(ii) the existence of the critical endpoint, and given the 
pressure-temperature ($P$-$T$) phase diagram \cite{Ref8:Lefebvre}. 
	Moreover, the ultrasonic study by Fournier \textit{et al.} has 
shown an anomaly related to 
(iii) the divergence of charge compressibility at the possible endpoint 
\cite{Ref9:Fournier}. 
	These three characters seem to support the criticality predicted by 
the dynamical mean field theory (DMFT) 
\cite{Ref10:Kotliar,Ref11:Kotliar,Ref12:Onoda}. 
	In order to confirm these qualitative features and go further to a 
substantial stage in the study of the Mott transition criticality, 
the electron transport measurements, which can detect the MIT directly, 
are highly desired.

	Quite recently, Limelette \textit{et al.} have reported transport 
measurements of $\kappa$-Cl under He gas pressure in Ref.14, where the 
hysteretic resistive anomaly associated with the Mott transition was observed 
and the possible endpoint was also supported. 
	However, the resistive transition was \textit{smooth} 
and the criticality on the endpoint was not clear, leading to the conclusion 
that the Mott transition in $\kappa$-Cl include a complex physics and can't 
be described simply by the DMFT \cite{Ref10:Kotliar}, which predicts that the Mott transition is 
in the same regime of the liquid-gas transition. 
	Thus, the criticality of the Mott transition is controversial. 
	In the present work, we have performed resistance measurements for 
$\kappa$-Cl crystals with high quality under the He gas pressure. 
	In contrast with the previous report \cite{Ref13:Limelette}, 
we observed the first-order Mott transition with huge 
\textit{discontinuous resistance jump} and 
\textit{sharp criticality} of the endpoint. 
	These behaviors, which are consistent with those of the liquid-gas 
transition in the Ising universality class as is suggested by the DMFT, give 
the experimental basis for the criticality of the Mott transition.

	The size of the $\kappa$-Cl crystal used here is 
$1.8 \times 1.4 \times 1.2$ mm$^{3}$. 
	The sample was mounted in the Be-Cu cell and pressurized by 
compressing the He gas. 
	The in-plane resistance was measured with the standard dc 
four-probe method under both isothermal pressure sweep and isobaric 
temperature sweep. 
	In the isothermal process, the pressure sweep was made so slowly that 
the temperature deviation was within $\pm$ 50 mK. 
	During the isobaric temperature sweep, which requires more care because 
the temperature sweep inevitably causes pressure change, the He gas inflow 
(outflow) to (from) the cell was finely controlled so that the pressure 
deviation was maintained within $\pm$ 0.05 MPa
	To ensure the hydrostatic nature of pressure, the present experiments 
were performed except the $P$-$T$ region of the He solidification. 
	Below $T_{\rm SC}$, the measurements were also made under a field of 
11 T normal to the conducting layer, which is much higher than the upper 
critical field, $H_{\rm C2}$, in the temperature range studied here.

\begin{figure}
\includegraphics[width=8.5cm,height=8.5cm]{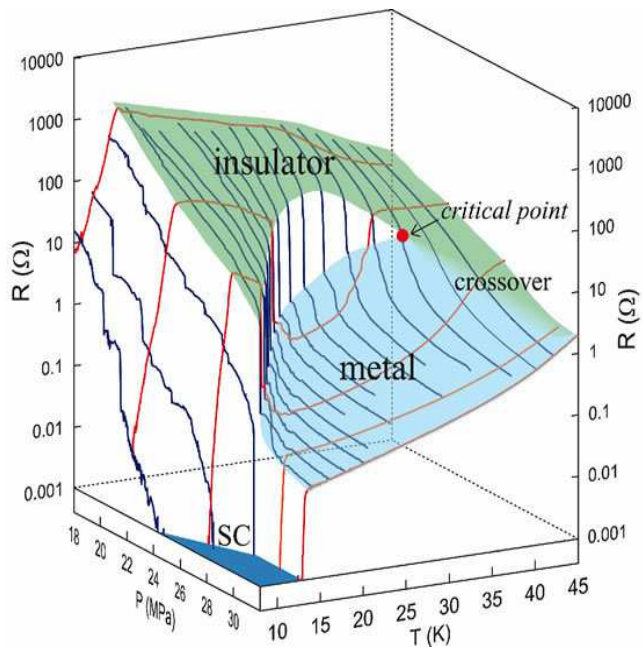}
\caption{\label{Fig1} Overview of resistance behavior around the Mott 
transition against pressure and temperature in $\kappa$-Cl. Red and blue 
lines are the data taken under isobaric temperature sweep and isothermal 
pressure sweep, respectively.}

\end{figure}

	The overall feature of the present results around the Mott transition 
is visualized in a pressure ($P$) - temperature ($T$) - resistance ($R$) 
diagram shown in Fig. \ref{Fig1}, where blue and red curves are data taken 
under the isothermal pressure sweep and the isobaric temperature sweep, 
respectively. 
	At low pressures, the system is highly resistive with non-metallic 
temperature dependence ($\partial R/\partial T < 0$), 
while at high pressures it is conductive with metallic temperature 
dependence ($\partial R/\partial T > 0$). 
	The transition between the two regimes occurs with a huge 
discontinuous resistance jump on a well-defined line in the $P$-$T$ plane. 
	It is also seen that the resistive jump becomes diminished at 
elevated temperatures and eventually vanishes at a certain critical point, 
above which the resistance variation is continuous. 
	Replacing the label of $z$-axis, $R$ ($\Omega$), in Fig. \ref{Fig1} 
by volume, $V$, of the classical liquid-gas system, one can see intuitively 
that the present $P$-$T$-$R$ diagram of the Mott transition has 
correspondence to the textbook $P$-$T$-$V$ diagram of the liquid-gas 
transition. 
	At low temperatures below 13 K, superconductivity appears at a high 
pressure region and even in the low pressure side resistance decrease with 
temperature is observed. 
	The data of the pressure dependence are classified into three 
temperature regions of $T > 38$ K, 38 K $> T >$ 13 K ($\sim T_{\rm SC}$), 
and $T <$ 13 K, which are discussed in detail below.

\begin{figure}
\includegraphics[width=8.5cm,height=10cm]{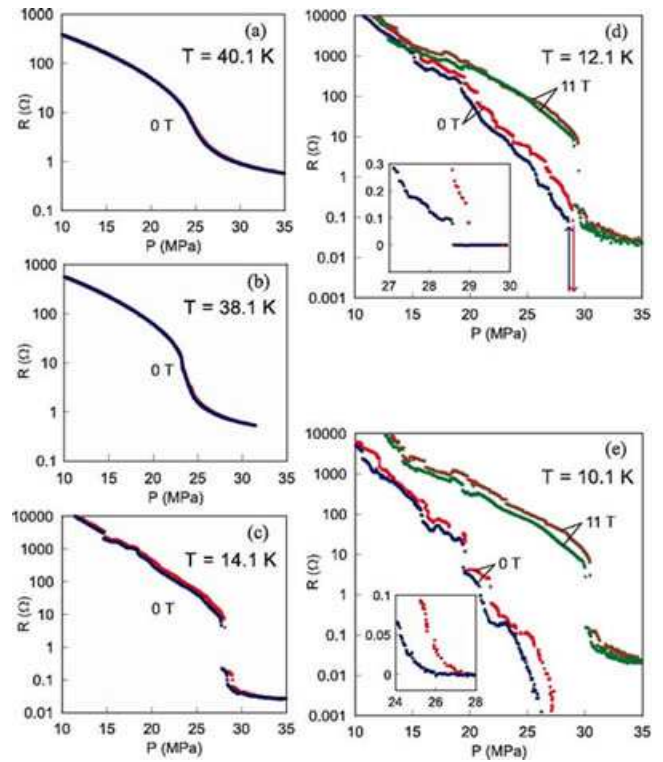}
\caption{\label{Fig2} Isothermal pressure dependence of resistance at various 
temperatures. 
	The red and blue points are the data taken under ascending and 
descending pressures, respectively, at a zero field ((a) - (e)), 
while the brown and green points are under ascending and descending pressures 
at a filed of 11 T ((d) and (e)). 
	Insets of (d) and (e) are enlarged views of each main panel in 
linear scales around the pressure where the resistance vanishes under a 
zero field. }

\end{figure}

\begin{figure}
\includegraphics[width=8.5cm,height=8.5cm]{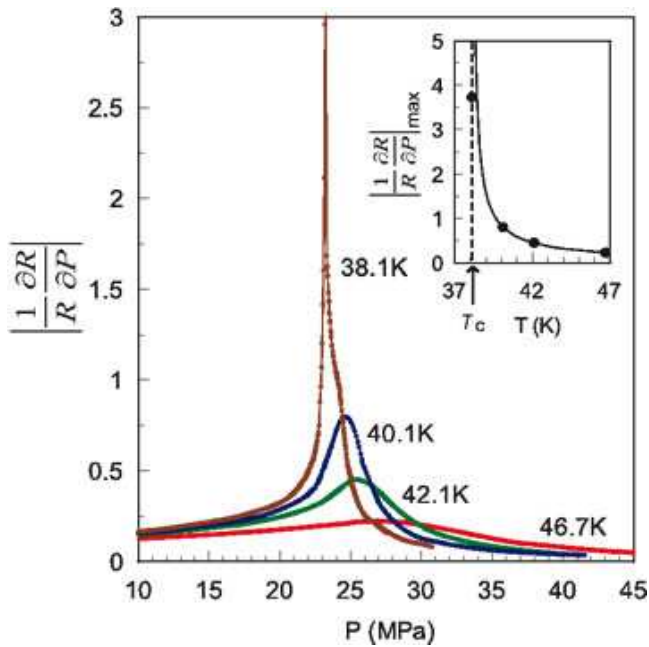}
\caption{\label{Fig3} The pressure derivative of resistance, 
$|\frac{1}{R}\frac{\partial R}{\partial P}|$, against pressure at several 
temperatures. 
	The value of $|\frac{1}{R}\frac{\partial R}{\partial P}|_{\rm max}$ is 
plotted against temperature in the inset. 
	The arrow indicates the critical temperature, 38.1 K, determined by 
disappearance of the resistance jump. 
The divergence is described by the solid curve of 
$\sim (T- T_{\rm {C}})^{- 0.9}$ with $T_{\rm {C}}$ of 38.1 K.}

\end{figure}

	As an example of the behaviors at higher temperatures above 38 K, 
the data at 40.1 K are shown in Fig. \ref{Fig2} (a), where neither anomaly 
nor hysteresis is observed. 
	Since the temperature derivative of resistance, 
$\partial R/\partial T$, is changed from negative to positive by pressure 
(see Fig. \ref{Fig1}), the resistance variation against pressure is regarded 
as crossover from insulator to metal. 
	As seen in Fig. \ref{Fig2} (b), the variation gets steeper with 
temperature decreased. 
	One can define a crossover pressure at which the pressure derivative 
of resistance, $|\frac{1}{R}\frac{\partial R}{\partial P}|$, shows a peak as 
shown in Fig. \ref{Fig3}. 
	The peak grows as temperature approaches 38.1 K, where it is 
divergent (see the inset of Fig. \ref{Fig3}). 
%
%
%
%
%
%
%
%
%
%
%
	Namely, the $|\frac{1}{R}\frac{\partial R}{\partial P}|$ is divergent 
around the critical point of ($P_{\rm C}$, $T_{\rm C}$) = (23.2 MPa, 38.1 K) 
roughly as $\sim (T - T_{\rm C})^{-(0.9 \pm 0.1)}$ against temperature along 
the crossover line. 
	This transport criticality observed here should be related to 
divergence in the charge compressibility suggested experimentally 
\cite{Ref9:Fournier} and predicted theoretically 
\cite{Ref11:Kotliar,Ref12:Onoda}. 
	As for the liquid-gas transition, the compressibility, 
$|\frac{1}{V}\frac{\partial V}{\partial P}|$, is divergent at the critical 
point. 
	The phenomenological correspondence implies the possibility that 
the Mott transition belongs to the same universality class with the 
liquid-gas transition, namely the Ising universality class. 
	The exponent which we extracted above corresponds to so-called 
'$\gamma$' in the scaling law. 
	Although the use of $|\frac{1}{R}\frac{\partial R}{\partial P}|$ in 
order to extract '$\gamma$' in the Mott transition leaves room to 
be considered, the tentative value of 0.9 $\pm$ 0.1 is close to 
$\gamma \sim 1$ in the mean field theory and $\gamma \sim 1.25$ in the 3D 
Ising model.

	Below 38 K, the resistive crossover is changed into the resistive 
transition of the first-order, which is evidenced by hysteresis and the 
resistance jump. 
	The observation of the clear resistance jump means not only the 
first-order nature but also the consistency with the DMFT, which predicts the 
criticality relevant to the vanishing of discontinuous nature around the 
endpoint like the discontinuity of the volume, $\Delta V$, at the liquid-gas 
transition. 
	Shown in Fig. \ref{Fig2} (c) are the data at 14.1 K, where the 
magnitude of the jump at 28.1 MPa amounts to nearly two orders of magnitude 
and a small hysteresis of $\sim$ 0.3 MPa lager than the experimental error 
of $\sim$ 0.1 MPa is appreciable. 
	The huge resistive jump indicates a bulky transition. 
	Additional small jumps and slightly irreversible resistance traces 
extended around the bulk transition are likely to come from inhomogeneous 
internal pressure in the sample, although the possibility of intrinsic phase 
separation with only tiny fraction of the secondary phase is not ruled out. 
	As temperature is increased, the magnitude of the resistive jump 
decreases with the hysteresis diminished and eventually vanishes at the 
critical point, (23.2 MPa, 38.1 K) \cite{Ref15:Comment}, where the first-order 
MIT changes to the crossover. 
	We suppose that the absence of the resistance jump and sharp 
criticality in the previous report \cite{Ref13:Limelette} is ascribable to the 
effect of disorder, which can make the simple nature inherent in the Mott 
transition sophisticated.

\begin{figure}
\includegraphics[width=7cm,height=7cm]{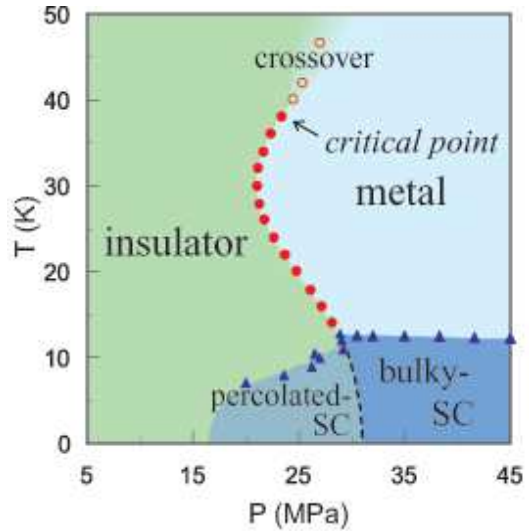}
\caption{\label{Fig4} Pressure-temperature phase diagram of $\kappa$-Cl. 
	Closed red circles and open red circles represent points at which 
resistance shows jump (first-order transition) and 
$|\frac{1}{R}\frac{\partial R}{\partial P}|$ is maximum (crossover point), 
respectively. 
	The superconducting transition defined by the resistance vanishing is 
marked by closed blue triangles. 
	The broken line is taken after Lefebvre \textit{et al.} [8] as 
the bulky SIT boundary.}

\end{figure}

	At lower temperatures below 13 K, superconductivity, which is a 
specific phase to the electronic systems, appears under pressure. 
	The data at 12.1 K under a zero field and 11 T are shown in 
Fig. \ref{Fig2} (d). 
	At a zero field, the resistance suddenly vanishes around 
28.5 - 29.0 MPa with large hyseteresis against pressure, indicating the 
bulky SC-insulator transition (SIT) of the first-order. 
	By application of 11 T, the SIT was converted into MIT with the 
transition pressure nearly unchanged. 
	It is seen that the field dependence of resistance is large even in 
the low pressure region below 28.5 MPa. 
	It is considered that tiny SC domains are induced progressively in 
the insulating host phase by pressure at a zero field but are destroyed at 
11 T \cite{Ref16:Comment}. 
	At 10.1 K (Fig. \ref{Fig2} (e)), the resistance under a zero field 
decreases continuously and falls below the noise level around $\sim$ 27 MPa, 
where the growing SC fraction is considered to be percolated before 
occurrence of the bulky SIT at a higher pressure. 
	Actually a bulky MIT under 11 T was at a higher pressure, 30.4 MPa, 
as seen in Fig. \ref{Fig2} (e). 
	This is consistent with the previous work by NMR \cite{Ref8:Lefebvre}, 
which shows coexistence of AFI and SC in a wide pressure range and drastic 
exchange of the volume fractions of AFI and SC phases at a certain pressure 
(broken lines shown in Fig. \ref{Fig4}).

	Figure \ref{Fig4} shows the $P$-$T$ phase diagram of $\kappa$-Cl, 
where the closed and open red circles represent points giving the resistance 
jumps (first-order transition) and maximum 
$|\frac{1}{R}\frac{\partial R}{\partial P}|$ (crossover point), respectively, 
and the superconducting transition defined by the resistance vanishing is 
marked by closed blue triangles. 
	Figure \ref{Fig4} is consistent with the Lefebvre's diagram 
\cite{Ref8:Lefebvre}, after which the broken line is drawn as a bulk SIT line.

\begin{figure}
\includegraphics[width=8cm,height=7cm]{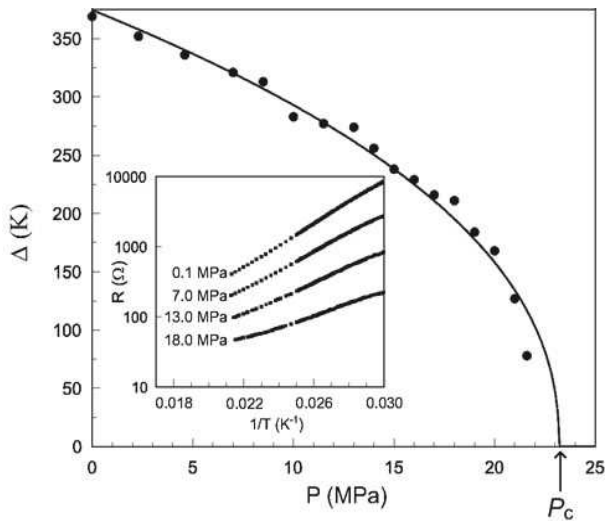}
\caption{\label{Fig5} Pressure-dependence of the effective charge gap. 
	The gap, $\Delta$, defined by $R \sim \exp (\Delta /T)$ is estimated 
from the Arrhenius plot of resistance as shown in the inset. 
	The arrow indicates the critical pressure, 23.2 MPa, determined by 
disappearance of the resistance jump. 
	The pressure dependence of the gap is described by the solid curve 
of $\sim (P_{\rm{C}} - P)^{0.4}$ with $P_{\rm{C}}$ of 23.2 MPa.}

\end{figure}

	We now examine the charge gap profile on the insulating side near the 
critical point. 
	The inset of Fig. \ref{Fig5} shows the Arrhenius plot of the 
resistance. 
	In a restricted temperature range from 50 K to 33 K, the data are 
nearly on straight lines, the slope of which gives the activation energy. 
	The effective charge gap, $\Delta$, defined by 
$R \sim \exp (\Delta /T)$ is shown against pressure in the main panel of 
Fig. \ref{Fig5}. 
	It is seen that the gap reasonably closes around the critical point. 
	The overall profile of the gap closing seems to be described by a 
form of $\Delta (P) \sim (P_{\rm C} - P)^{ 0.4 \pm 0.1}$ with $P_{\rm C}$ = 
23.2 MPa. 
	This is an additional criticality of the Mott transition and unique 
to the electronic systems. 
	In the context of the DMFT, Kotliar predicted that the density of 
states at the Fermi energy, $\rho (0)$, shows continuous change against 
temperature and the change is sharpened at the endpoint \cite{Ref10:Kotliar}. 
	The similar behavior is also expected against pressure. 
	The criticality of the charge gap can be related to the growth of 
$\rho (0)$ and its neighboring profile. 

	To conclude, the Mott transition in the quasi-two-dimensional organic 
system, $\kappa$-(BEDT-TTF)$_{2}$Cu[N(CN)$_{2}$]Cl, is accompanied by the 
gigantic resistive jump and the first-order transition line has an endpoint 
at ($P_{\rm C}, T_{\rm C}$) = (23.2 MPa, 38.1 K), where the transition is 
changed to the crossover. 
	The endpoint has the following critical behaviors; 
(i) the pressure derivative of resistance, 
$|\frac{1}{R}\frac{\partial R}{\partial P}|$, is divergent, reflecting the 
divergence in the charge compressibility, 
(ii) the resistance jump vanishes, and 
(iii) the effective charge gap closes. 
	The criticalities, (i) and (ii), establish the phenomenological 
correspondence between the Mott transition of correlated electrons and the 
liquid-gas transition of molecules or atoms as predicted by the dynamical mean 
field theory. 
	The present results provide the experimental basis for physics of the Mott transition criticality.

\begin{acknowledgments}
	The authors acknowledge N. Nagaosa, S. Onoda, and M. Imada for fruitful discussion.
\end{acknowledgments}


\begin{thebibliography}{99}
\bibitem{Ref15:Comment} Even at 38.1 K, a small kink, which is the remnant of 
the jump, is appreciable. The actual critical temperature may be slightly 
higher than 38.1 K.
\bibitem{Ref16:Comment} The resistance decrease with temperature below 13 K 
at a zero field was changed into non-metallic temperature dependence by 
application of 11 T.
\end{thebibliography}
\end{document}